\documentclass[twocolumn]{article}

\usepackage[a4paper]{geometry}
\usepackage{verbatim}
\usepackage{soul}
\usepackage{url}

\usepackage{graphicx}	
\usepackage{amsmath}	
\usepackage{amssymb}	
\usepackage{algorithm2e} 
\usepackage{natbib}
\usepackage{algorithmic}
\usepackage{booktabs}
\usepackage{multirow}
\usepackage{color}
\usepackage{ulem}
\usepackage{authblk}
\graphicspath{ {./}{images/} } 

\def\apjs{\ref@jnl{ApJS}}               
\def\aap{\ref@jnl{A\&A}}                
\def\mnras{\ref@jnl{MNRAS}}             

\SetAlCapSkip{0.5em}

\title{Accelerating Dedispersion using Many-Core Architectures} 

\author[1,2]{Jan Novotn\'y}
\author[1]{Karel Ad\'amek}
\author[4]{M.~A. Clark}
\author[3]{Mike Giles}
\author[1]{Wesley Armour}

\affil[1]{Oxford e-Research Centre, Department of Engineering Science, University of Oxford, 7 Keble Road, Oxford OX1 3QG, UK}
\affil[2]{Research Centre for Theoretical Physics and Astrophysics, Institute of Physics, Silesian University in Opava, Czech Republic}
\affil[3]{Mathematical Institute, University of Oxford, Oxford, OX2 6GG, UK}
\affil[4]{NVIDIA, 2788 San Tomas Expressway, Santa Clara, CA 95051, USA}
\date{}

\begin{document}
\maketitle

\begin{abstract}
Astrophysical radio signals are excellent probes of extreme physical processes that emit them. However, to reach Earth, electromagnetic radiation passes through the ionised interstellar medium (ISM), introducing a frequency-dependent time delay (dispersion) to the emitted signal. Removing dispersion enables searches for transient signals like Fast Radio Bursts (FRB) or repeating signals from isolated pulsars or those in orbit around other compact objects. The sheer volume and high resolution of data that next generation radio telescopes will produce require High-Performance Computing (HPC) solutions and algorithms to be used in time-domain data processing pipelines to extract scientifically valuable results in real-time. This paper presents a state-of-the-art implementation of brute force incoherent dedispersion on NVIDIA GPUs, and on Intel and AMD CPUs. We show that our implementation is 4x faster (8-bit 8192 channels input) than other available solutions and demonstrate, using 11 existing telescopes, that our implementation is at least 20 faster than real-time. This work is part of the AstroAccelerate package.
\end{abstract}

\section{\label{sec:intro}Introduction}
An upcoming new generation of radio telescopes, such as the Square Kilometre Array (SKA) \citep{carilli:2004:ska, ska}, will simultaneously observe many different regions of the radio sky. Each simultaneous observation will have high time resolution and fine channelisation of the observed bandwidth, giving rise to large data volumes at high data rates. These data are expected to make storing all data for offline analysis impractical, necessitating faster than real-time data processing software. 

To extract the very faint signals present in the noisy data produced by these telescopes, many processing steps have to be performed on the data. One of the more computationally expensive steps is dedispersion. The dedispersion process increases the signal-to-noise ratio (SNR) of received signals from the emitting object being studied by shifting samples in different frequency channels in time, thus correcting for the time delay introduced by dispersion. Samples at the same time-stamp are then summed over frequency channels, increasing the SNR and probability of detection.

The dispersion of the emitted pulse occurs due to the interaction between photons in the pulse and the ionised interstellar medium (ISM) through which they travel. Dispersion has the effect of causing a frequency dependent time delay ($\Delta \tau$) in the photon's propagation. Specifically lower frequency photons within the pulse $(f_\mathrm{low})$ are observed later than their high frequency $(f_\mathrm{high})$ counterparts \citep[see][]{Lorimer:2005}. This time delay is proportional to the inverse square of the frequency, given by the cold plasma dispersion law
\begin{equation} 
    \Delta\tau = \mathrm{DM}~C_\mathrm{DM} \biggl(\frac{1}{f^{2}_\mathrm{low}} - \frac{1}{f^{2}_\mathrm{high}} \biggr), \label{eq:qcp}
\end{equation} 
with the constant of proportionality $C_\mathrm{DM} = 4148.8\times 10^{3}$\,MHz$^{2}$\,pc$^{-1}$\,cm$^{3}$\,s. The parameter DM in Equation~\eqref{eq:qcp} is referred to as the dispersion measure and is defined as the integral of the electron column density ($n_\mathrm{e}$) along the line of sight (distance $d$) between source and observer, i.e.
\begin{equation} 
    \mathrm{DM} = \int_{0}^{d} n_\mathrm{e}\,\mathrm{d}l\,. \label{eq:dm}
\end{equation}

Given the quadratic relationship between time delay and frequency (see Equation~\eqref{eq:qcp}), dispersion is governed by a~single free parameter, the dispersion measure. When searching for new events from unknown sources, the distance between the object and the observer is unknown, which means all possible dispersion measures must be calculated and searched. Detecting and studying such events in real-time on the scale required for the next generation of radio telescopes, together with the computational complexity of dedispersion, necessitates a high performance computing (HPC) solution for real-time observation and detection~\citep{Armour:2012:adass_ddtr, Fluke:2012, Barsdell:2010}.

To remove the effects of dispersion two different approaches can be used, coherent and incoherent dedispersion. The coherent approach uses information about the observed phase of the pulse to reconstruct the pulse profile as it was emitted (within the limits of the in-homogeneous scattering of the ISM). The incoherent method applies the appropriate time delay to each independent frequency channel in channelised intensity data. Although the coherent method is more accurate and has higher sensitivity, computational requirements are far more demanding than for the incoherent method. 
As such, when performing surveys of the radio sky, it is common to employ incoherent dedispersion.

Several different codes exist with differing implementations of dedispersion specifically for Graphics Processing Units (GPUs) \citep{Magro:2011, Barsdell:2012, Sclocco:2016, Bassa:2017, Zackay:2017, Kong:2021:dedispersion}. There are other implementations of the dedispersion transform for example using Fast Fourier transforms by~\citet{Bassa:2022:FourierDedispersion}. This approach is however not capable of detecting FRBs or accelerated pulsars due to no or weak Fourier response to these signals. In this article we present our implementations of dedispersion for different computer architectures, NVIDIA GPUs, Intel and AMD CPUs. We compare the performance of these implementations with the state of the art packages, specifically looking at data processing rates and sensitivity.  

All of the implementations that we present in this article have been developed for AstroAccelerate\footnote{\url{https://github.com/AstroAccelerateOrg/}} \citep{AstroAccelerate, Dimoudi:2018, adamek2020:SPDT, Novotny:2022:streams, Adamek2020:JERK, adamek:2022:harmonic, white:2023:bits}, a many-core accelerated software package for processing time domain radio astronomy data. AstroAccelerate is actively used as part of scientific pipelines like MeerTRAP \citep{MeerTRAP2021_a, MeerTRAP2021_b}, Greenburst \citep{GREENBURST2020}.

Incoherent dedispersion is described in Section \ref{sec:DirectTransform}, whilst the implementation for the CPU is presented in Section~\ref{sec:implementation_cpu} and GPU in Section~\ref{sec:implementation_gpu}. In Section~\ref{sec:results} we discuss performance results of AstroAccelerate for different scenarios on selected many-core platforms and compare these results with the performance of other software packages like Heimdall. Real-time performance on selected telescopes is presented in Section~\ref{sec:performance_telescope} and the conclusion are summarised in Section~\ref{sec:discussion}.
\section{\label{sec:DirectTransform}The Direct Dispersion Transform}
Incoherent dedispersion is the process of shifting detected power data in time inside each individual frequency channel which collectively makes up the total telescope bandwidth. Shifts are applied to counter the effect of interstellar (or intergalactic) dispersion before integrating the data over the frequency bandwidth of the telescope to increase the signal-to-noise ratio of astrophysical signals detected by the telescope.

Here we present the direct (sometimes called brute force) approach to de-dispersing measured power data. As well as being the simplest approach for performing the task of dedispersion, it has two significant advantages. The first is that the algorithm is exact, by this we mean that the errors associated with this approach are at the discretization level of the instrument. The second is that the algorithm can be written in such a way that is particularly suited to execution on many-core devices. 

\begin{figure}[htb!]
	\begin{center}
		\includegraphics[width=\linewidth]{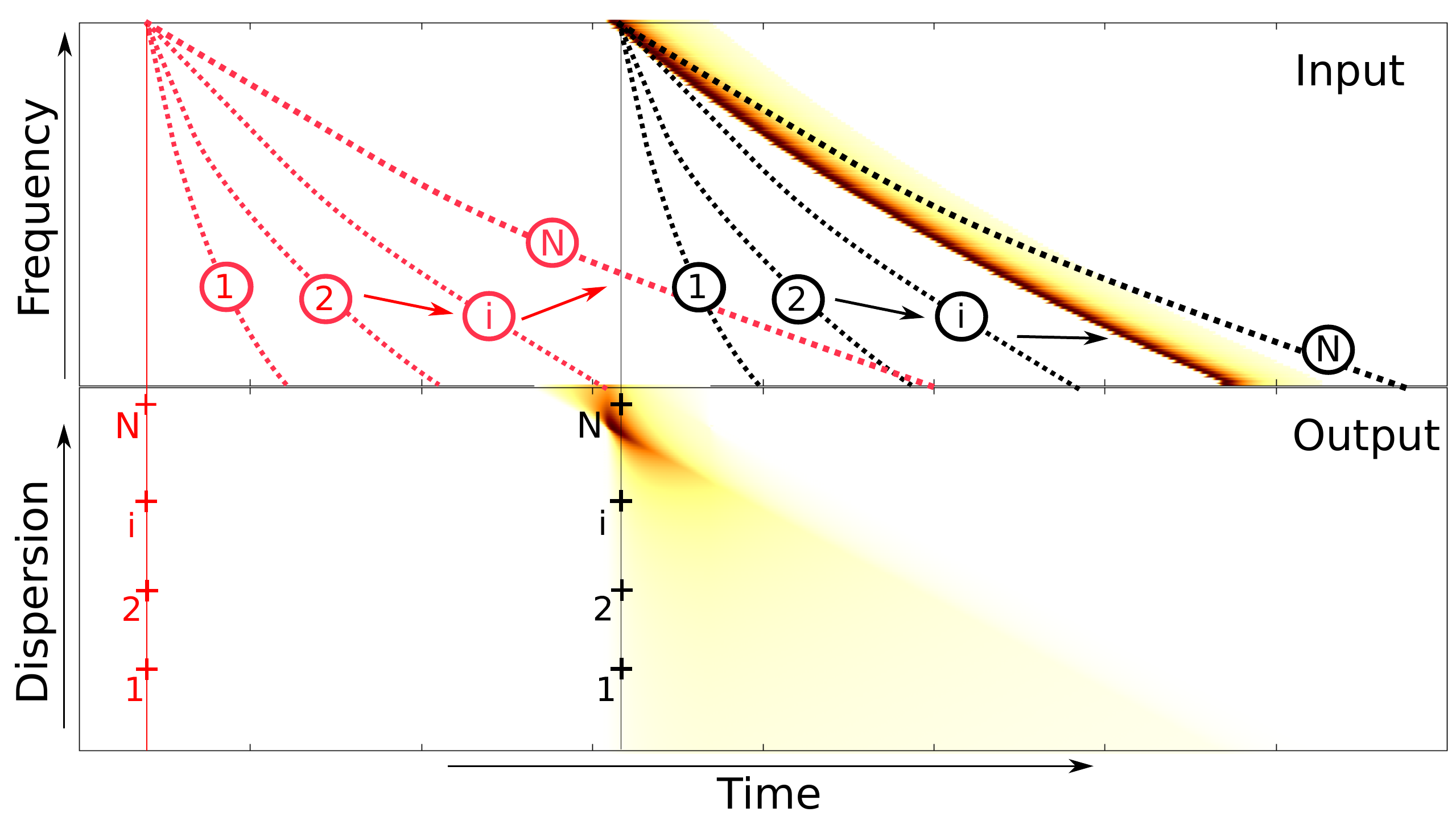}
		\caption{\label{fig:explain_ddtr}Representation of the incoherent dedispersion approach. Top represents the input data and the bottom the results (output). For simplicity we present a clear single signal. The dotted lines correspond to individual DM trials (the summation of data points along the line) computed for a given time sample (showed for DM trials 1, 2, i  and N). The sum of the line maps to one point (cross) in the output field.}
	\end{center}
\end{figure}

Incoherent dedispersion can be algebraically expressed as 
\begin{equation} 
    S(DM,t) = \sum_{f=0}^{N_{f}} x(f,t+\Delta t(DM,t,f)) \label{eq:algebra} 
\end{equation}
where a frequency dependent shift in time $\Delta t(DM,t,f)$ is calculated for each digitised frequency channel. By substituting $f_\mathrm{low} = f_\mathrm{c} - \Delta f/2$ and $f_\mathrm{high} = f_\mathrm{c} + \Delta f/2$ into Equation~\eqref{eq:qcp} we can express the time shift in the form
\begin{equation} 
    \Delta t(DM,t,f) \approx 8297.4 \times 10^{3} \biggl(\frac{DM \Delta f}{f_\mathrm{c}^{3}}\biggr) 
\end{equation}
where $f_\mathrm{c}$ corresponds to the central frequency of the band and $\Delta f$ is a finite bandwidth that is $\Delta f \ll f_\mathrm{c}$. 
Applying the correct time shifts to each frequency channel results in a shifted signal that appears as though it has arrived at the same instant in time. The process of incoherent dedispersion is shown pictorially in Figure~\ref{fig:explain_ddtr} in which dotted lines (red and black) correspond to different DM trials $1,2,...,N$. From Equation~\eqref{eq:algebra} we can derive a simple pseudo-code (see Algorithm \ref{al:direct}) that outlines the direct dedispersion approach.

\begin{algorithm}[htb!]
    \caption{\label{al:direct}Pseudo-code for the direct dispersion transform, where $N_\mathrm{t}$ is the number of time samples, $N_\mathrm{f}$ is the number of dispersion measures searched and $N_\mathrm{f}$ is the number of frequency channels.} 
    \KwData{$x(f,t)$}
    \KwResult{$DM(dm,t)$}
        \For{$t=0,\ldots,N_\mathrm{t}-1$}{
        	\For{$dm=0,\ldots,N_\mathrm{DM}-1$}{
        		$S(dm,t)=0.0$\;
        		\For{$f=0,\ldots,N_\mathrm{f}-1$}{
        			$h=\Delta t(dm,t,f)$\\
        			$S(dm,t) \gets S(dm,t) + x(f,t+h)$\\
			}
			$DM(dm,t) \gets S(dm,t)$\\
        	}
	   }
        \KwRet $DM(dm,t)$
\end{algorithm}

From pseudo-code~\ref{al:direct} we see that for $N_{DM}$ trial dedispersion searches over power data that has $N_{t}$ time samples and $N_{f}$ frequency samples the computational complexity of the algorithm is $\mathcal{O}(N_{DM} N_{t} N_{f})$.

The arithmetic intensity $I$ is another important characteristic of an algorithm. The roofline model \cite{Williams:2009:roofline} defines $I$ as a ratio of the number of floating point operations performed by the algorithm per amount of data required in bytes read or written by the algorithm to RAM or GPU main memory in the case of GPUs. The value of $I$ can help us to identify what will limit the performance of an algorithm. When the algorithm is limited by the number of floating point operations it needs to perform, it is called a compute-bound algorithm. If the memory bandwidth limits the algorithm by not providing enough data per second, we have a memory-bound problem.

In order to decide whether an algorithm on a given hardware platform (CPU, GPU) is compute-bound or memory-bound, we need to look at the critical arithmetic intensity $I_\mathrm{crit}$ that represents a turning point from an algorithm being memory-bound to being compute-bound and vice versa. For a given hardware platform, the critical arithmetic intensity $I_\mathrm{crit}$ is a ratio of computational performance in FLOPS and the memory bandwidth in bytes. If $I_\mathrm{alg}$ of an algorithm is $I_\mathrm{alg}<I_\mathrm{crit}$ that algorithm is memory-bound. For $I_\mathrm{alg}>I_\mathrm{crit}$, the algorithm will be compute-bound. On the modern hardware $I_\mathrm{crit}>1$, see Table \ref{tab:hardware} for values of $I_\mathrm{crit}$.

The dedispersion's arithmetic intensity for a single DM trial is given as:
\begin{equation}
    I_\mathrm{d} = \frac{n_\mathrm{o}}{n_\mathrm{b}}=\frac{N_\mathrm{f}-1}{N_\mathrm{f} + 4} \rightarrow 1\,,
    \label{dedispOI}
\end{equation}
where $n_\mathrm{o}$ is the number of floating point operations performed, $N_\mathrm{f}$ is the number of frequency channels and $n_\mathrm{b}$ is the number of bytes required. We have assumed that incoming data are 8-bit and the output data from the dedispersion is FP32 (4 bytes). Thus the dedispersion transform will be memory-bound on the most modern hardware platforms. Therefore data reuse in an available cache must be utilised to increase dedispersion performance.

\section{CPU Implementation}
\label{sec:implementation_cpu}
Our CPU implementation of the incoherent dedispersion algorithm is written in the C programming language with OpenMP and Cilk Plus, where OpenMP is used for parallelisation across cores on multi-core CPUs and Cilk Plus is used to express fine grained parallelism and allows the compiler to effectively vectorize parts of the code.

\begin{algorithm}[h]
\caption{\label{alg:cpu2}Pseudo-code of the parallel CPU direct dedispersion transform.}
    \KwData{$x(f,t)$}
    \KwResult{$DM$}
\emph{\#{}pragma omp parallel for collapse(2)}\\
		\For{$i=0$ to $N_\mathrm{t}$; by $\mathrm{D}_\mathrm{t}$}{
			\For{$jj=0$ to $N_\mathrm{DM}$ by $\mathrm{D}_\mathrm{dm}$}{
				\For{$kk=0$ to $N_\mathrm{f}$ by $\mathrm{D}_\mathrm{f}$}{
						\For{$j=jj$ to $(jj+\mathrm{D}_\mathrm{dm})$; $j{+}{+}$}{						
							int $S$[$\mathrm{D}_\mathrm{t}$]\;
							\eIf{kk = 0}{ $S$[:] = 0\;}
							{$S$[:] = $DM$[$j\times N_\mathrm{t}+i{:}\mathrm{D}_\mathrm{t}$]\;}
							\For{$k=kk$ to $(kk+\mathrm{D}_\mathrm{f})$; $k{+}{+}$}{
								$h$=$\Delta t(j,i,k)$\;
								$S$[:] += $x$[$k\times N_t + i + h{:}\mathrm{D}_\mathrm{t}$]\;
							}
							$DM$[$j\times N_\mathrm{t}+i{:}\mathrm{D}_\mathrm{t}$] = $S$[:]\;
						}
				}
			}
		}
		$DM$[0:$N_{DM}\times N_\mathrm{t}$] /= channels\;
		\KwRet DM
\end{algorithm}

\begin{figure}[htb!]
    \centering
    \includegraphics[width=\linewidth]{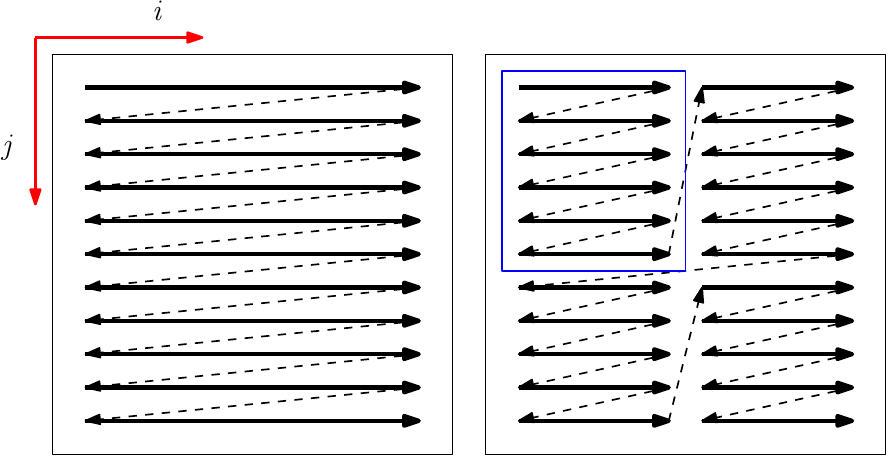}
    \caption{\label{fig:cpu_tiling_basic}Schematic example of tiling optimisation technique in case of two dimensions ($i$ and $j$). On the left and on the right we see the situation when loop blocking is not used and the case when it is. The original large array is partitioned into smaller blocks (blue rectangle), which can fit into cache size.}
\end{figure}

\begin{figure}[htb!]
    \centering
        \includegraphics[width=\linewidth]{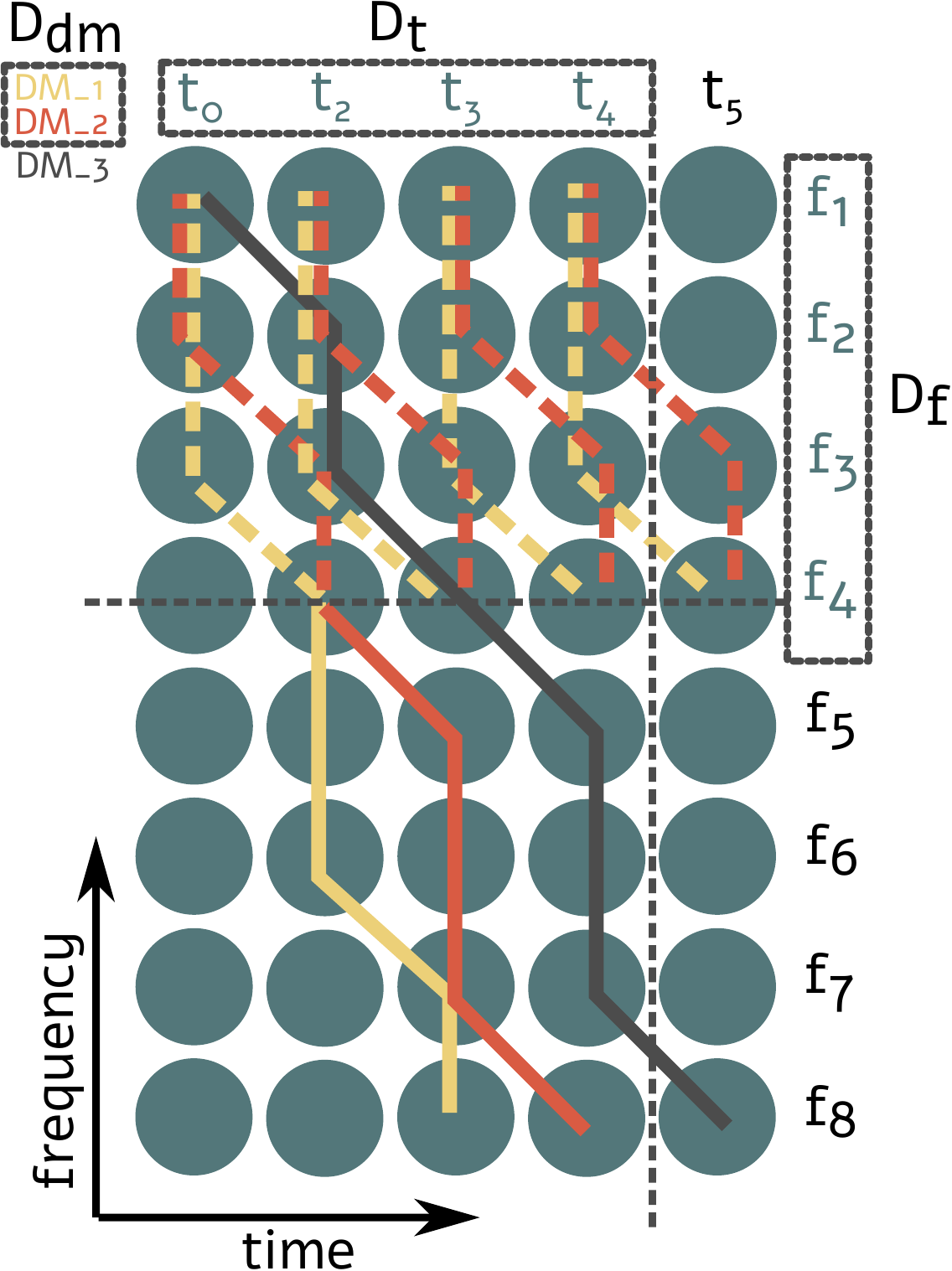}
    \caption{\label{fig:cpu_tiling_astro} Example of the loop tiling in the case of the CPU dedispersion algorithm. One thread computes a partial sum of $D_\mathrm{f}$ frequency channels ($f_1$--$f_4$) of $D_\mathrm{dm}$ for a number of DM trials ($DM_1$ and $DM_2$) and for $D_t$ time samples with the same DM trials ($t_0$--$t_4$); all of these are represented by the coloured dashed lines, where different colours correspond to different DM trials, the $x$-axis the time samples and $y$-axis the frequency channels.}
\end{figure}

As discussed in the previous section dedispersion is a memory bandwidth bound algorithm. Hence to achieve good performance careful use of CPU cache is required. As such, our CPU algorithm implements a very well known optimisation technique called loop tiling (see Figure~\ref{fig:cpu_tiling_basic}), also known as loop blocking, or strip mine and interchange \citep{Wolf:1991:loop_tiling}, which transforms the input domain (memory) to smaller chunks able to fit into cache, thereby improving the locality of data accesses in loops. This technique also helps reduce the number of cache misses.\footnote{When working with caches it is important to optimise for cache hits. In principle, a cache hit occurs when the cache contains data needed by a thread otherwise a cache miss is generated, meaning that data has to be reloaded from main memory, thus reducing performance.} Specifically we use ``tiling'' in frequency channels as well as in dedispersion trials (DMs), where each thread calculates: 1) several DM-trials for neighbouring time samples (achieving good spatial locality in cache), 2) several neighbouring DM-trials (achieving good temporal locality in cache\footnote{The amount of reuse is dependent on the closeness of neighbouring DM-trials.}, as a cacheline can be used multiple times to update multiple DM-trials). A schematic overview of our partitioning of the data space is shown in Figure~\ref{fig:cpu_tiling_astro}. 

In the pseudo-code (see Algorithm~\ref{alg:cpu2}) the size of the tiles are represented by $D_\mathrm{f}$ for frequency channels and $D_\mathrm{dm}$ in the case of the DM.
For the code to achieve the best performance, optimal values of $D_\mathrm{f}$, $D_\mathrm{dm}$ need to be found as well as the optimal number of time samples per thread ($D_\mathrm{t}$). These optimal values are dependent not only on the CPU used, but also on the input telescope data parameters (like central frequency, number of frequency channels, etc.) and the DM survey plan.

\section{GPU implementation}
\label{sec:implementation_gpu}%
For a GPU code limited by the memory bandwidth to the GPU main memory, it is essential to reuse data and effectively and efficiently use the L1/L2 cache or the user-managed cache called the shared memory. For peak performance, we need to ensure three things. Firstly, the accumulator that stores the integrated value of the frequency channels $S_\mathrm{loc}$ must be stored in the fastest area of memory available. Secondly, the data for each frequency channel that will undergo the dedispersion transform must be readily available to the GPU's streaming multiprocessors; a compute unit analogous to CPU cores. Finally, to avoid costly evaluation of the power-law by each thread, the value of the dedispersion shift should be calculated using as few operations as possible.

The advantage of the shared memory over an L1/L2 cache is that the user can control data locality. The shortfall of the shared memory is its size. Where the L1 cache can defer to the larger but slower L2 cache, the shared memory has no such option. Any implementation that uses shared memory and relies on a custom data structure will be limited by its size. This limitation gives rise to the two different algorithms outlined below. In short, the shared memory version of the direct dedispersion transform can process most shift values with high performance; a cache version, while slower, can handle even large shifts, often present at lower central frequencies. 

\begin{figure}[htb!]
    \centering
    \includegraphics[width=0.7\linewidth]{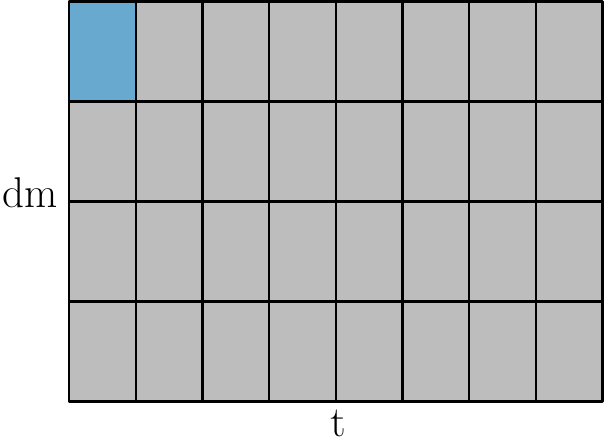}
    \caption{The output (dm,~t) plane is partitioned into sections of size ($D_\mathrm{dm}$,~$D_\mathrm{t}$) that is processed by a single thread-block, shown in blue. } \label{fig:dmt}
\end{figure}

\begin{figure}[htb!] 
    \begin{center} 
    \includegraphics[width=\linewidth]{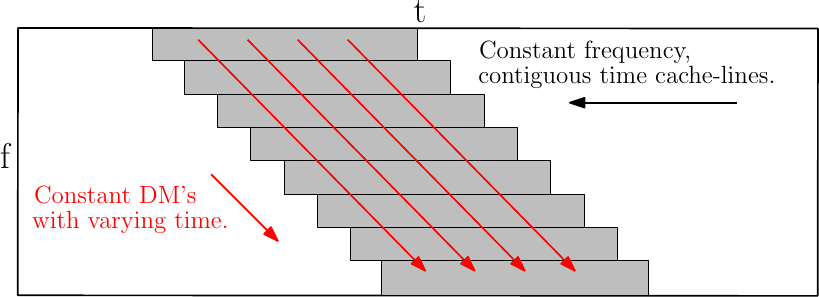}
    \caption{\label{fig:cachelines} Data are loaded from cache lines of constant frequency, and contiguous time (grey boxes). Multiple dedispersion trials DM$=\mathrm{constant},t$ values (red lines) are held in registers and each thread updates a set of these in parallel.}
    \end{center}
\end{figure}

Both of our GPU algorithms are written in the CUDA C/C++ programming language and use the following methodology: One GPU thread processes several time elements for a fixed value of dispersion. A thread stores accumulated values into registers\footnote{Registers are the fastest area of GPU memory. However, increasing the number of registers that a GPU kernel uses reduces the number of resident thread-blocks that can occupy each SM.}. Nearby values of time and DM in the output $\mathrm{DM(dm,t)}$ space are grouped together into thread-blocks (fig. \ref{fig:dmt}) such that a single thread-block calculates $D_\mathrm{t}$ time samples and $D_\mathrm{dm}$ dispersion steps. The size of the area processed by a single thread-block is tunable. Data from the input $x(f,t)$ are read in a coalesced manner ensuring the best possible performance. 

The higher performing shared memory version is described below. The cache version will not be described further. In the results section the cache version is only used where we cannot use the shared memory version of our code.

\subsection{Shared Memory Algorithm}

The pseudo-code for the GPU kernel is presented in Algorithm \ref{al:gpu-shared}. In the shared memory GPU implementation of the dedispersion algorithm, each thread-block calculates a different part of the output $\mathrm{DM(dm,t)}$ space as shown in Figure~\ref{fig:dmt}. A single thread from a thread-block loops over frequency channels with step of $D_\mathrm{ch}$ channels, where it loads a single element of the $x(f,t)$ data into a local buffer $B_\mathrm{loc}$ using the shift
\begin{equation}
\Delta t_\mathrm{bl} = T_\mathrm{bl}\Delta t_\mathrm{pc}\,, \label{eq:part_shift}
\end{equation}

where the coefficient $T_\mathrm{bl}$ in Eq.~\eqref{eq:part_shift} represent the lowest DM calculated by the thread-block, and $\Delta t_\mathrm{pc}$ are the coefficients of the cold plasma dispersion law for each frequency channel. All threads in the thread-block thus form a contiguous (in time) block of $x(f,t)$ data in shared memory. This is shown in Figure~\ref{fig:cachelines}.

When local buffer $B_\mathrm{loc}$ is loaded, each thread sums appropriate elements in shared memory using the differential shift $\Delta t_\mathrm{diff} = T_\mathrm{diff}\Delta t_\mathrm{pc}$ and updates their value of the partial sum $S_\mathrm{loc}$, held in registers, that in the end will result into dedispersed value expressed by Eq.~\eqref{eq:algebra}. The coefficient $T_\mathrm{diff}$ represents DM values calculated by different threads within the range of DM values calculated by a thread-block ($D_\mathrm{dm}$).
After this, threads integrate the loop over frequency channels and a new block of $x(f,t)$ data is loaded.

To avoid costly evaluation of the power-law by each thread the calculation of the dispersion shift $\Delta t$ was split into two parts. The first part is an array of pre-calculated coefficients of the cold plasma dispersion law $\Delta t_\mathrm{pc}$ which are evaluated as:
\begin{equation}
    \Delta t_\mathrm{pc}(i) = 4148.74\left(\frac{1}{\left(f_\mathrm{high} - \Delta f \cdot i\right)^2} - \frac{1}{f_\mathrm{high}^2} \right)\,,
\end{equation}

where $\Delta t_\mathrm{pc}(i)$ represent time shift for $i$-th frequency channel, $f_\mathrm{high}$ is the highest frequency of the telescope bandwidth and $\Delta f$ is a bandwidth of single frequency channel. This array is then scaled in the thread-block by the DM value that is being evaluated. The second part required for calculation of $\Delta t$ are the two DM coefficients: $T_\mathrm{bl}$ which is constant within the thread-block, and $T_\mathrm{diff}$ which relates to DM calculated by a thread (see Algorithm~\ref{al:gpu-shared}).

Further performance improvements can be achieved by processing multiple time samples $N_\mathrm{REG}$ per thread. This exploits instruction level parallelism, where a thread can process multiple independent instructions. The adverse effect is that processing too many time samples per thread increases register usage too much, leading to fewer active threads and lower performance.

When working with input data of 8-bits or less, we pack the data into 32-bit words and then use integer addition to achieve SIMD (single instruction, multiple data) in word. In the case of 8-bit data, this allows us to process two time samples per operation, increasing throughput significantly. 

The values of $N_\mathrm{REG}, D_\mathrm{t}$, $D_\mathrm{dm}$ and $D_\mathrm{ch}$ have a significant impact on the performance of this code and must be tuned for a given telescope configuration and DM plan to gain maximum performance.

\begin{algorithm} 
\caption{\label{al:gpu-shared}Pseudo-code (GPU kernel) for the shared memory based GPU algorithm.}
    \SetKw{KwBy}{by}
    \SetKwFunction{shared}{\_\_shared\_\_}
    \SetKwFunction{synch}{\_\_synchthreads()}
    \SetKwFunction{threadx}{ThreadIdx.x}
    \SetKwFunction{thready}{ThreadIdx.y}
    \SetKwFunction{blockx}{BlockIdx.x}
    \SetKwFunction{blocky}{BlockIdx.y}
    \SetKwFunction{dedisp}{Dedispers}
    \SetKwData{thx}{thx}
    \SetKwData{thy}{thy}
    \KwData{$x(f,t)$, $\mathrm{DM}_\mathrm{start}$, $\mathrm{DM}_\mathrm{step}$, $\Delta t_\mathrm{pc}$}
    \KwResult{$\mathrm{DM(dm,t)}$}
    \emph{Initiate local accumulator}\;
    $S_\mathrm{loc}[N_\mathrm{REG}]=0$\;
    \emph{Shared memory buffer to store local copy of x(f,t)}\;
    \shared $B_\mathrm{loc}(f,t)$\;
    \emph{Time shift depends on position in $\mathrm{DM(dm,t)}$ plane}\;
    $T_\mathrm{diff} = \thready \times \mathrm{DM}_\mathrm{step}$\;
    $T_\mathrm{bl} = \mathrm{DM}_\mathrm{start} + \blocky\times D_\mathrm{dm} \times \mathrm{DM}_\mathrm{step}$\;
    
    \For{$c=0$ \KwTo $\mathrm{N}_{ch}$ \KwBy $D_\mathrm{ch}$}{
        \emph{Data segment is stored into shared memory}\;
        $B_\mathrm{loc}(f,t)$ = $x(f,t + T_\mathrm{bl}\Delta t_\mathrm{pc}(c))$\;
        \synch\;
        \For{$l=0$ \KwTo $D_\mathrm{ch}$}{
            \emph{Dedisperse local data into accumulators}\;
            $S_\mathrm{loc}=$\dedisp$(B_\mathrm{loc}(f,t+T_\mathrm{diff}\Delta t_\mathrm{pc}(l)))$\;
        }
    }
    \emph{Store local results into output $\mathrm{DM(dm,t)}$}\;
    $\mathrm{DM(dm,t)} = S_\mathrm{loc}$\;
    \KwRet $\mathrm{DM(dm,t)}$\;
\end{algorithm}

\section{\label{sec:results}Results}
In this section, we first explore how the performance of our GPU dedispersion code, which is part of the AstroAccelerate package, depends on the parameters of input data and the dedispersion plan. To compare our dedispersion to other implementations, we have considered two test cases. The first test measures the execution time of the direct dedispersion transform for a varying number of frequency channels using a fixed dedispersion plan (subsection~\ref{sec:test1}). The second test, described in subsection~\ref{sec:test2}, demonstrates the performance of our dedispersion implementation when used in a data processing pipeline. For this test, we have used three datasets with different central frequencies and dedispersion plans. We compare the output from the different implementations tested in subsection~\ref{sec:output_comparison}. Finally, we demonstrate AstroAccelerate performance on selected radio telescopes in subsection~\ref{sec:performance_telescope}.

We compare our results with currently existing and in use implementations of direct dedispersion suited for HPC environments. Specifically, for the first test and the numerical difference test, we use the code ``Dedisp''~\citep{Barsdell:2012} and ``Dedispersion''~\citep{Sclocco:2016}. Due to similar names we will refer to these dedispersion implementations by associated processing pipelines. The Heimdall\footnote{\url{https://sourceforge.net/projects/heimdall-astro/}} pipeline is a GPU accelerated transient detection pipeline that utilises ``Dedisp'' while ``Dedispersion'' is used in the Amber pipeline \citep{amberpipeline}. 

The Amber dedispersion code is written using the Open Computing Language (OpenCL) programming language, while Heimdall's Dedisp uses the CUDA C/C++ programming model. Both implementations can run in two modes/regimes. Heimdall, in ``adaptive'' mode, changes the DM step during dedispersion depending on the parameter of DM tolerance set by the user. The number of DM trials thus varies. The second mode of Heimdall can be described as ``fixed'', meaning that the user selects a fixed step size in the DM range, thus the pipeline outputs a fixed number of DM trials. In the case of Amber, for single DM, the frequency channels are divided into sub-bands which are firstly dedispersed (``step one''), and then dedispersed within each sub-band (``step two''). Our code performs a fixed number of DM trials with a fixed DM step for each DM range akin to the ``fixed'' mode in Heimdall and one of the mode in Amber.

In all following tests, the input signal is generated using ``fake'' from the pulsar processing package SIGPROC. We generated synthetic filterbank files for 4-bit, 8-bit and 16-bit precisions.

We have used two GPU cards (NVIDIA Tesla V100~-- Volta generation, NVIDIA A100~-- Ampere generation), two Intel processors (Xeon Phi 7290~-- Knights Landing (KNL), Xeon Gold 6230~-- Cascade Lake) and AMD processor EPYC 7601~-- Naples. Their hardware specifications can be found in the Table~\ref{tab:hardware} where the GPU shared memory bandwidth is calculated as
\begin{eqnarray}
    \mathrm{BW (bytes/s)} = \mathrm{(bank\, bandwidth\, (bytes))} \times \nonumber\\ 
    \times \mathrm{(clock\, frequency\, (Hz))} \times \nonumber\\
    \times \mathrm{(32\, banks)} \times \mathrm{(\#\, multiprocessors)}\,,
\end{eqnarray}
and the CPUs theoretical peak performance by
\begin{eqnarray}
    \mathrm{Peak\,(GFlops)} = \mathrm{(clock\,frequency (GHz))} \times \nonumber \\
    \mathrm{(\#\,of\,Cores)} \times \mathrm{(vector\,width)} \times\nonumber\\
    \times \mathrm{(instructions/cycle)} \times \mathrm{(FLOPs/instruction)}\,,
\end{eqnarray}
where FLOPs/instruction=2 in the case of fused multiply-add (FMA), and the vector width is 16 for single precision and 8 for double precision. Intel CPUs (Xeon Gold 6230, Xeon Phi 7290) have two AVX-512 units thus instructions/cycle=2, for AMD EPYC 7601 is instructions/cycle=1.  In the case when all cores utilise AVX-512 instructions the core clock frequency is reduced by 100--200\,GHz \citep{intel-scalable, intel-scalable-xeon}.

For the time measurements of Heimdall and AstroAccelerate, we used the NVIDIA Compute profiler software (ncu) for the GPUs and \verb+omp_get_wtime+ function for the CPU case, whilst in the case of Amber, we followed the supplied timer using OpenCL functions. Where possible, we use the supplied auto-tuning scripts for each test and all codes to achieve the best performance. Unless otherwise stated, all execution times show kernel run-time only, that is, no data transfers from host to device (GPU) are taken into account. The GPU codes are compiled with \verb+nvcc+ compiler and the CPU codes with the Intel compiler (\verb+ICC+). Codes based on OpenCL are compiled with \verb+nvcc+ for GPUs and \verb+ICC+ for CPUs using appropriate OpenCL flag enabled. Used compiler flags are summarised in~Table~\ref{tab:hardware}.

\begin{table*}[htb!]
	\caption{\label{tab:hardware}Hardware specifications and compiler specifications on the tested GPUs and CPUs. The value of the memory bandwidth in brackets on Xeon Phi 7290 corresponds to the case of using the 16\,GB Multi-Channel DRAM (MCDRAM).}
\begin{center}
\resizebox{\linewidth}{!}{
\begin{tabular}{@{}lrrlrrr@{}}
\toprule
 &
  \multicolumn{1}{l}{V100 SXM2} &
  \multicolumn{1}{l|}{A100 SXM4} &
   &
  \multicolumn{1}{l}{Xeon Gold 6230} &
  \multicolumn{1}{l}{Xeon Phi 7290} &
  \multicolumn{1}{l}{AMD EPYC 7601} \\ \midrule
CUDA cores            & 5120     & \multicolumn{1}{r|}{6912}      & \# of Cores/Threads   & 20/40       & 72/288        & 32/64    \\
\# of SMs             & 80       & \multicolumn{1}{r|}{108}       & Base Clock Frequency  & 2.1 GHz     & 1.5 GHz       & 2.2 GHz  \\
Boost Core Clock      & 1455 MHz & \multicolumn{1}{r|}{1410 MHz}  & Frequency for AVX-512 & 2.0 GHz     & 1.3 GHz       & ---      \\
Memory Clock          & 877 MHz  & \multicolumn{1}{r|}{1215 MHz}  & Mem. bandwidth        & 140.8 GB/s  & 115.2 (400+) GB/s  & 276.573 GB/s\\
Dv. m. bandwidth      & 900 GB/s & \multicolumn{1}{r|}{1555 GB/s} & Cache size            & 27.5 MB L3  & 36 MB L2      & 64 MB L3 \\
Shared m. bandwidth &
  14899 GB/s &
  \multicolumn{1}{r|}{19492 GB/s} &
  DP compute &
  --- &
  --- &
  1126 GFLOPS \\
Memory size           & 32768 MiB    & \multicolumn{1}{r|}{40960 MiB}     & DP compute (AVX-512)  & 1280 GFLOPS & 2995.2 GFLOPS & ---      \\
TDP                   & 300 W    & \multicolumn{1}{r|}{400 W}     & TDP                   & 125 W       & 245 W         & 180 W    \\
Critical Arithmetic Intensity & 15.4 &\multicolumn{1}{r|}{11.6} & Critical Arithmetic Intensity & 9 & 7.5 & 4 \\ \bottomrule
Other specifications: &
  \multicolumn{1}{l}{} &
  \multicolumn{1}{l}{} &
   &
  \multicolumn{1}{l}{} &
  \multicolumn{1}{l}{} &
  \multicolumn{1}{l}{} \\ \midrule
NVIDIA driver:        & \multicolumn{6}{r}{495.29.05}                                                                              \\
CUDA version:         & \multicolumn{6}{r}{11.5.119}                                                                               \\
ICC version:          & \multicolumn{6}{r}{18.0.3}                                                                                 \\
OpenCL version (CPU): & \multicolumn{6}{r}{Intel 2019.3.208}\\
\bottomrule Compiler flags: &
  \multicolumn{1}{l}{} &
  \multicolumn{1}{l}{} &
   &
  \multicolumn{1}{l}{} &
  \multicolumn{1}{l}{} &
  \multicolumn{1}{l}{} \\ \midrule
\multirow{2}{*}{AMD EPYC 7601:} & \multicolumn{6}{r}{-std=c99 -O2 -Wall -Wextra -qopenmp -march=core-avx2 -qopt-prefetch}\\ 
& \multicolumn{6}{r}{-fma -ftz -fomit-frame-pointer -finline-functions -qopt-streaming-stores=never}\\
\multirow{2}{*}{Xeon Phi 7290:} & \multicolumn{6}{r}{-qopenmp -fp-model fast=2 -std=c99 -O2 -fma -xMIC-AVX512 -align}\\ 
& \multicolumn{6}{r}{-finline-functions -no-prec-div -ipo -DOPEMP\_SPEC -qopt-streaming-stores=never}\\ 
\multirow{1}{*}{Xeon Gold 6230:} & \multicolumn{6}{r}{same as Xeon 7290 except: -march=core-avx2}\\
\bottomrule                                 
\end{tabular}
}
\end{center}
\end{table*}

\subsection{Performance dependency}
\begin{figure*}[ht!]
    \centering
    \includegraphics[width=.48\linewidth]{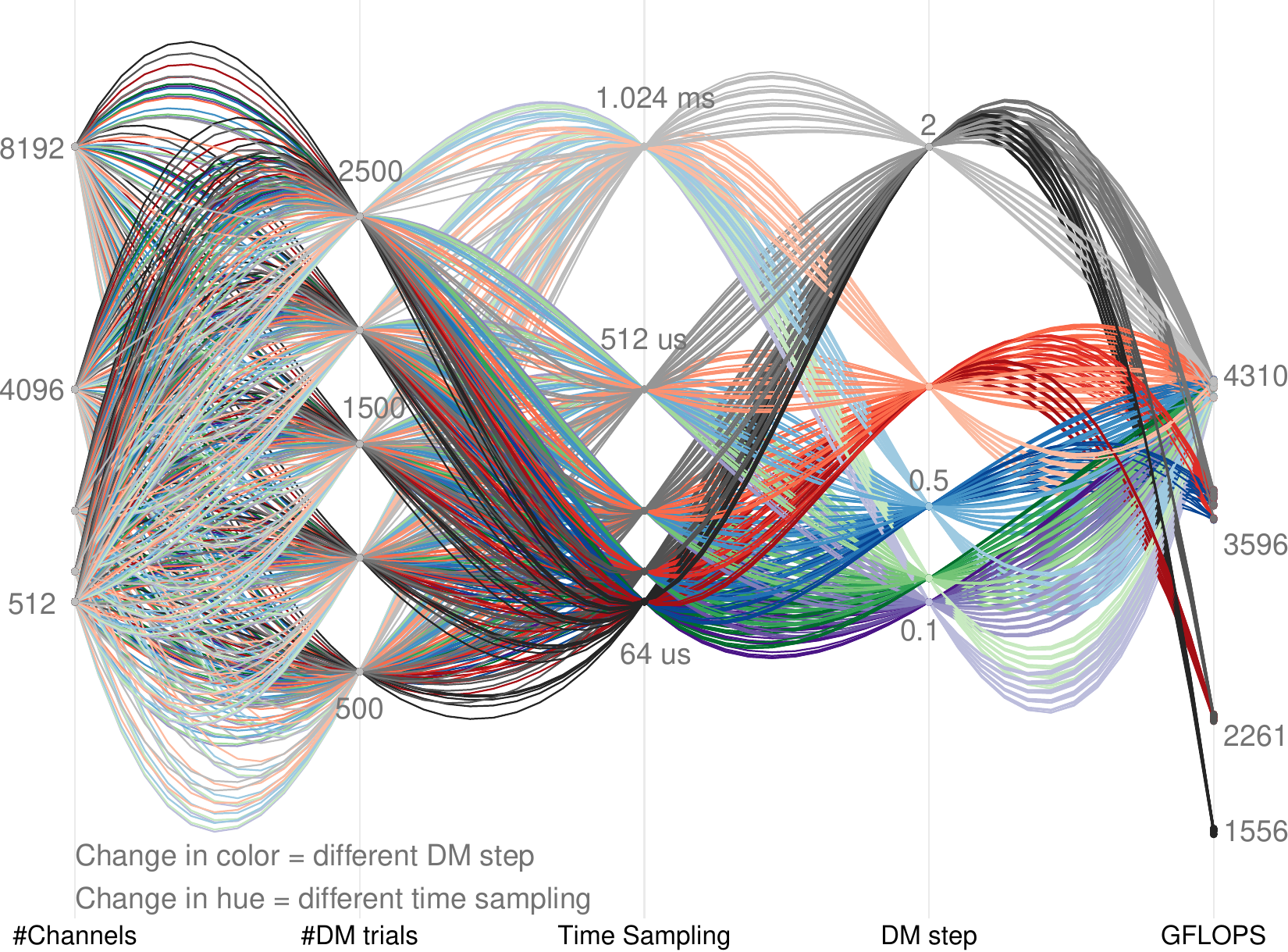}\hfill%
    \includegraphics[width=.48\linewidth]{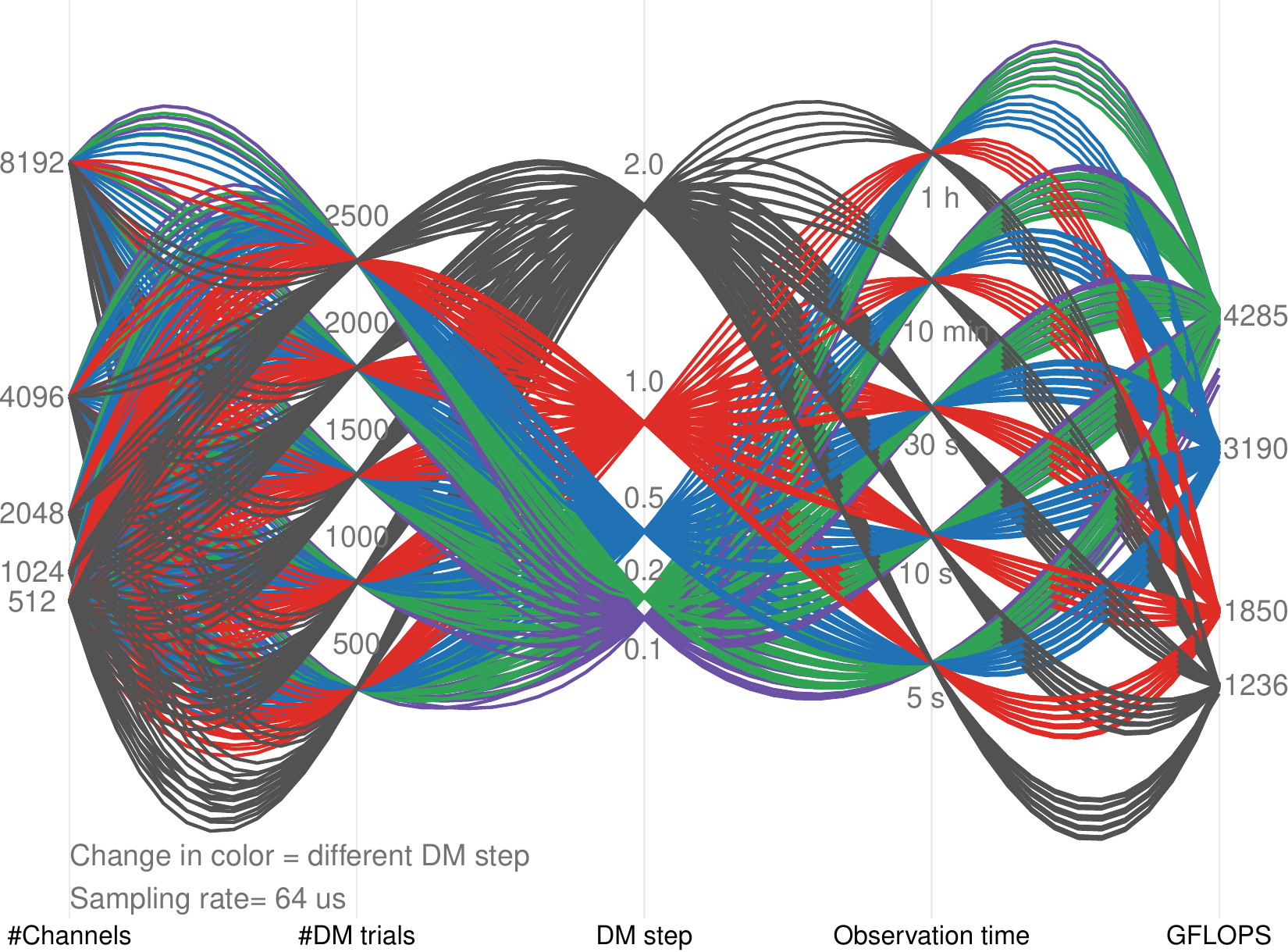}
    \caption{\label{fig:plan_survey}Flow-plot representing how AA performance (GFLOPS) depends on parameters of the data and dedispersion plan. Each line symbolises one run of the autotuned AA GPU dedispersion with data and a dedispersion plan described by parameters shown in the figure. These parameters are the number of channels, DM trials, DM step, samplings time of the data and observation time. For simplicity, only 8-bit input data and telescope central frequency set to 1400\,MHz are shown. The left plot, where we have fixed the observation time and emphasised the DM step (colour) and sampling time (hue), shows that the performance is mainly affected by the combination of two parameters, the DM step and the sampling time. Lines of rich dark colour, where high sampling time is combined with a large DM step, show the least performing case, where shared memory GPU code cannot be used. Lines of brighter colour have increasing performance as the sampling time gets lower. The right plot, where the sampling time is fixed and only the DM step is emphasised, shows that the performance is independent of the observation time.}
\end{figure*}

To illustrate the dependency of AstroAccelerate's performance on different input data parameters, we have visualised all distinct cases in the form of a flow-plot shown in Figure~\ref{fig:plan_survey}. The parameters used are the number of frequency channels, sampling time of the data, observation time, the number of DM trials and size of the DM step. The performance is expressed in GFLOPS as this can be understood as an average performance per second. The execution time will still depend on the size of the task. The left plot in Figure~\ref{fig:plan_survey} shows that the right combination of time sampling with DM step size is essential for high performance. The plot on the right of Figure~\ref{fig:plan_survey} shows that the performance of the  AA dedispersion code does not dependent on the observation time, as in all other tested codes.

\subsection{\label{sec:test1}Frequency resolution test}
The following test shows the behaviour of the execution time of all mentioned codes when we change the number of channels.

For testing, we created a synthetic signal with 4-bit, 8-bit and 16-bit samples with a time sampling of 64~$\mu$s for five different channelisations: 512, 1024, 2048, 4096 and 8192, where the last one corresponds to the maximum number of channels that Heimdall can safely process due to integer overflow. The length of the input signal corresponds to 10~s of observation data with a central frequency 1400\,MHz and a total bandwidth of 300\,MHz. This observation length is sufficient to get representative performance measurements. Moreover, it allows us to extend our testing up to 8192 channels also for the Heimdall dedispersion code, which cannot be run at this number of channels for longer observation times. For the survey plan, we use three DM ranges without time binning (also known as downsampling), which together search for signals ranging from a DM 0 to 500~pc\,cm$^{-3}$. We limited the search to 500~pc\,cm$^{-3}$ because at high DM it is common to use downsampling/scrunch factor which we do not include in this test. Moreover, the high DM searches are covered in the second test. The dedispersion plan used is summarised in Table~\ref{tab:plans}.

\begin{table}[htb!]
\begin{center}
    \caption{\label{tab:plans} Dedispersion plans used in frequency resolution test Sec.~\ref{sec:test1}.}
    \begin{tabular}{@{}rrr@{}}
        \toprule
        DM range & DM step & \multirow{2}{*}{\# DM trials} \\
        (pc\,cm$^{-3}$) & (pc\,cm$^{-3}$) & \\ \midrule
        0--150   & 0.10 & 1500 \\
        150--300 & 0.20 & 750 \\
        300--500 & 0.25 & 800 \\ \bottomrule
    \end{tabular}
\end{center}
\end{table}

\begin{figure}[htb!]
    \centering
    \includegraphics[width=.99\linewidth]{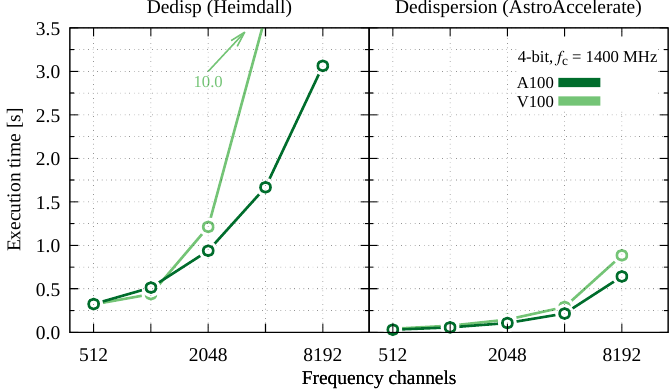}\\
    \includegraphics[width=.99\linewidth]{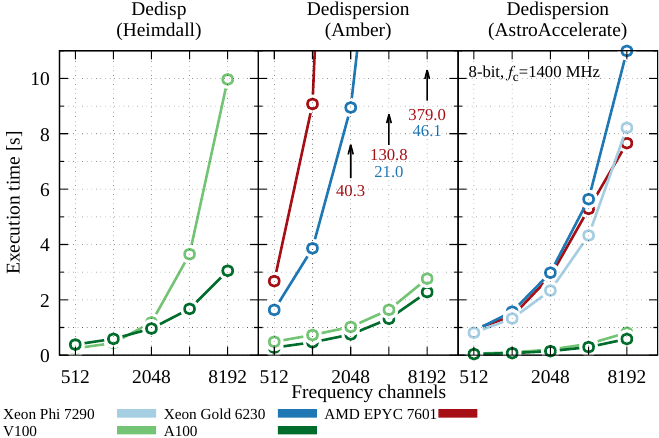}\\
    \includegraphics[width=.99\linewidth]{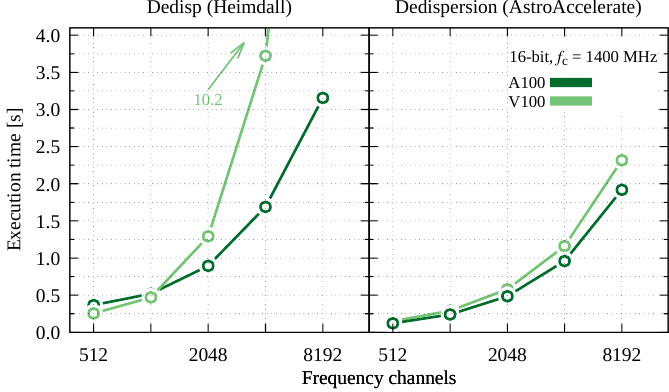}
    \caption{\label{fig:time-channels}The execution time of the corresponding dedispersion plan (see Table~\ref{tab:plans}) for 4-bit, 8-bit and 16-bit precision (from top to bottom) input data with an increasing number of channels, observation time $T=10\,s$, central frequency $f_\mathrm{c}=1400$\,MHz and total bandwidth of 300\,MHz.}
\end{figure}

\begin{figure}[htb!]
    \centering
    \includegraphics[width=.99\linewidth]{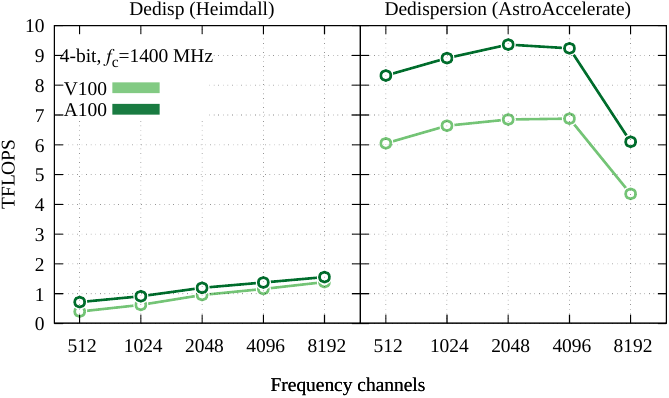}\\
    \includegraphics[width=.99\linewidth]{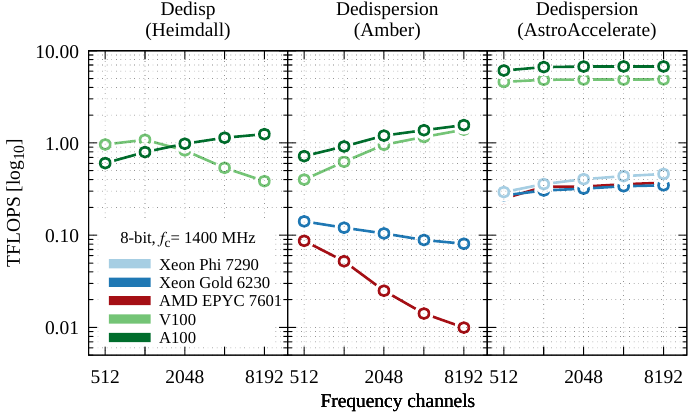}\\
    \includegraphics[width=.99\linewidth]{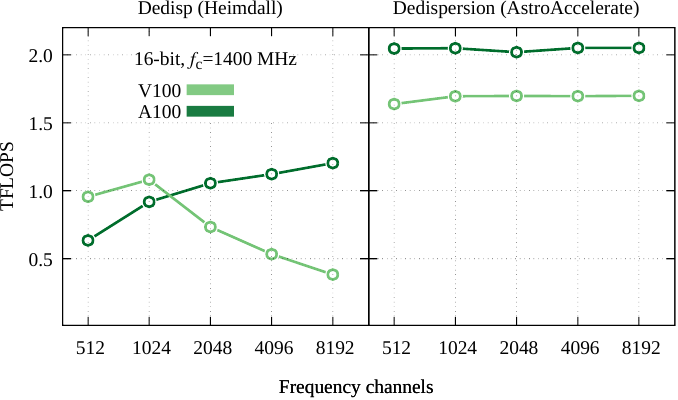}\\
    \caption{\label{fig:flops-channel}Performance in FLOPS with an increasing number of channels for all tested codes. The first three rows correspond to 4-bit, 8-bit and 16-bit precision.}
\end{figure}

The execution time of the dedispersion plan using a differing number of frequency channels and bit precisions for all tested codes are shown in Figure~\ref{fig:time-channels}. The results of Amber Dedispersion for the 4-bit and 16-bit are missing because the code did not return the correct results of the injected signal from SIGPROC. There are also no results for our CPU implementation because it only supports 8-bit precision. Although the OpenCL parallel language can be used across platforms, Intel officially does not support KNL. Therefore we do not show execution times for Amber.\footnote{Even though it is still possible to get OpenCL code running on KNL by using an older Intel OpenCL driver without the support of AVX-512~\citep{Johnston-Milthorpe:2018:Dwarfs}, i.e., it provides only half of the theoretical peak performance.} Figure \ref{fig:flops-channel} shows how performance, expressed in the floating point operations per second (FLOPS), changes with increasing numbers of channels.

\begin{figure}[htb!]
    \centering
    \includegraphics[width=\linewidth]{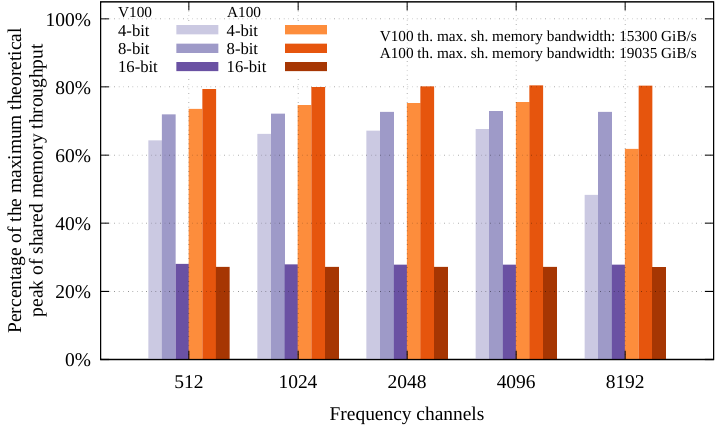}
    \caption{\label{fig:shared_throughput}Percentage of shared memory bandwidth to the theoretical maximum achieved by AstroAccelerate dedispersion for different bit precisions and an increasing number of channels. The theoretical maximum of shared memory bandwidth for each GPU card is in the top right corner.}
\end{figure}

By analysing our implementation of the incoherent dedispersion on the GPU using the NVIDIA Nsight Compute, we see that in the case of 4-bit and 8-bit precision our implementation is limited by the shared memory bandwidth on both GPU cards, whilst the 16-bit precision version is limited by the special function units ($\approx95\%$ of utilisation) also on both GPU cards. The special function units take care of type conversions that are required by the 16-bit implementation. The summary of shared memory bandwidth utilisation by AstroAccelerate for each card and bit precision is presented in Figure~\ref{fig:shared_throughput}.

\subsection{\label{sec:test2}Processing pipelines}
In this section, we analyse the execution time of the AstroAccelerate running the dedispersion plan with different DM ranges (0--3000\,pc\,cm$^{-3}$) and time binning factors (also known as downsampling/scrunch factors). We compare the results alongside the GPU accelerated pipeline -- Heimdall. We have not compared the Amber pipeline as this was compared to Heimdall by \citet{amberpipeline}.

As the radio telescopes operate on a wide range of central frequencies, we selected three scenarios to demonstrate the performance and behaviour of both AstroAccelerate and Heimdall. The selected central frequencies are $f_c=$400 (low), $f_c=$800 (mid) and $f_c=$1400 (high), each with typical bandwidth, sampling rate, number of channels, and DM surveys plans (for details see Table~\ref{tab:pipeline}). The synthetic input data were generated as in the previous cases using SIGPROC ``fake'' The observation lengths correspond to ${\sim}300$\,s for the low central frequency case, and ${\sim}50$\,s for the others.

\begin{table*}[htb!]
\caption{\label{tab:pipeline}Specifications of the input data used for pipeline comparison. The observation length corresponds ${\sim}$300\,s for the low central frequency and ${\sim}$50\,s for mid and high central frequencies.}
\begin{center}
\begin{tabular}{@{}lrrrrr@{}}
\toprule
 & \multicolumn{1}{c}{Central} & \multicolumn{1}{c}{Total} & Sampling & \# channels & \multicolumn{1}{c}{DM}\\
 & \multicolumn{1}{c}{frequency} &  \multicolumn{1}{c}{bandwidth} &  \multicolumn{1}{c}{rate} & & \multicolumn{1}{c}{range} \\\midrule
Low & 400 MHz & 200 MHz & 256 $\mu$s & 1024 & 0--1500\\
Mid & 800 MHz & 200 MHz & 128 $\mu$s & 4096 & 0--2000\\
High & 1400 MHz & 300 MHz & 64 $\mu$s & 4096 & 0--3000\\ \bottomrule
\end{tabular}
\end{center}
\end{table*}

\begin{figure}[htb!]
    \centering
    \includegraphics[width=\linewidth]{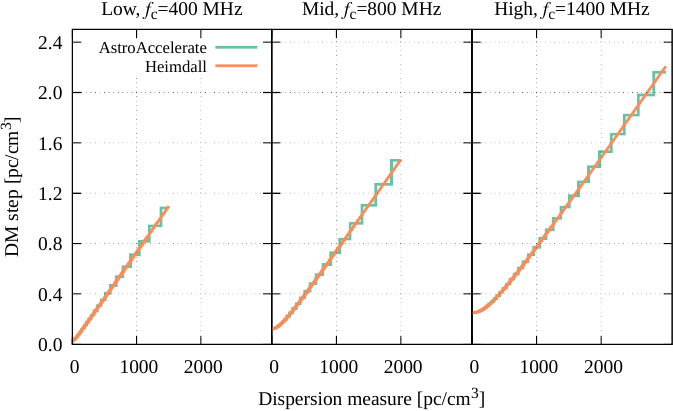}
    \caption{\label{fig:plan_comparison} Visualisation of the continuous Heimdall survey DM plan with a discrete AstroAccelerate DM plan for all three cases: $f_c=$400, 800 and 1400\,MHz (from left to right). Details can be found in Table~\ref{tab:pipeline}.}
\end{figure}

To compare both pipelines fairly and use all the implemented features, we must ensure they use the same or comparable DM plan. Heimdall, by default, uses an ``adaptive'' mode. That is, Heimdall generates a list of DMs during the start-up phase of the pipeline based on the smearing tolerance factor between individual DM trials (default value 1.25), thereby changing its DM step for each DM trial. Whereas AstroAccelerate uses a user-defined fixed DM step for each DM range. To obtain a similar DM plan for AstroAccelerate, we used the ``continuous'' plan from Heimdall and created the closest ``discrete'' plan for our pipeline. A visualisation of these plans for all three cases is shown in Figure~\ref{fig:plan_comparison}. Note that the highest DM is different for each case (${\sim}$1500, 2000 and 3000\,pc\,cm$^{-3}$).

\begin{figure*}[htb]
    \centering
    \includegraphics[width=.32\linewidth]{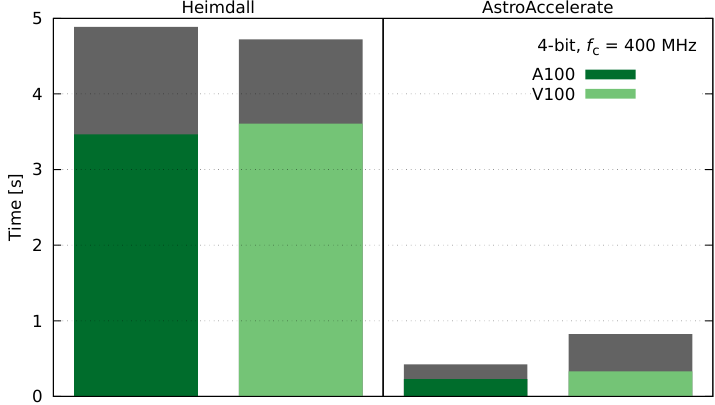}\hfill%
    \includegraphics[width=.32\linewidth]{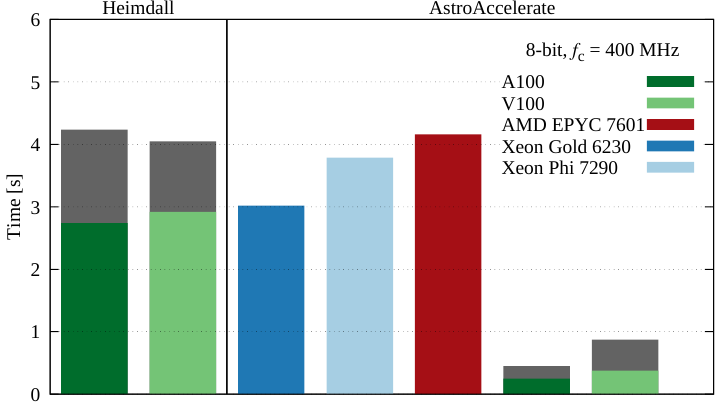}\hfill%
    \includegraphics[width=.32\linewidth]{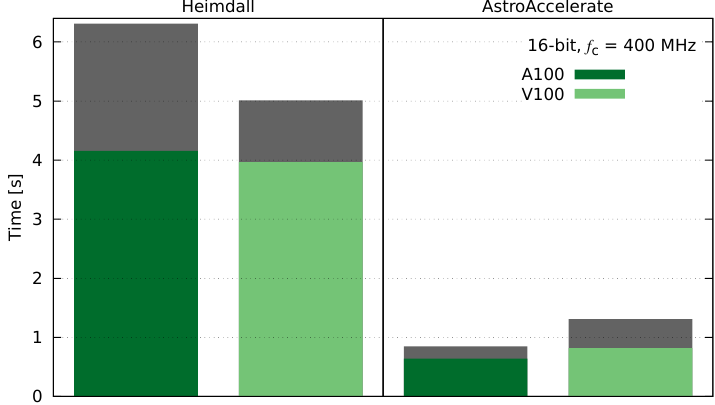}\\
    \includegraphics[width=.32\linewidth]{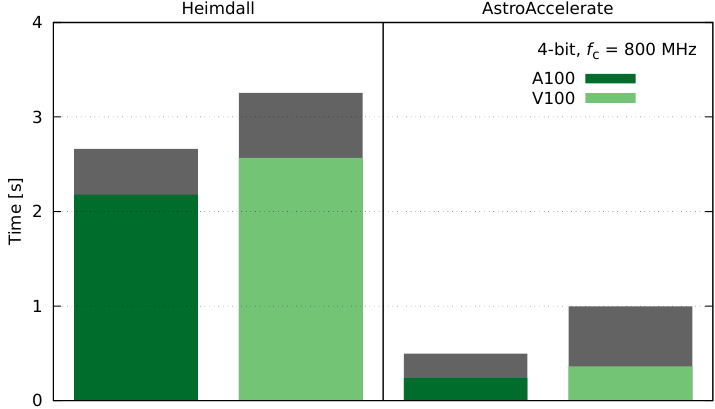}\hfill%
    \includegraphics[width=.32\linewidth]{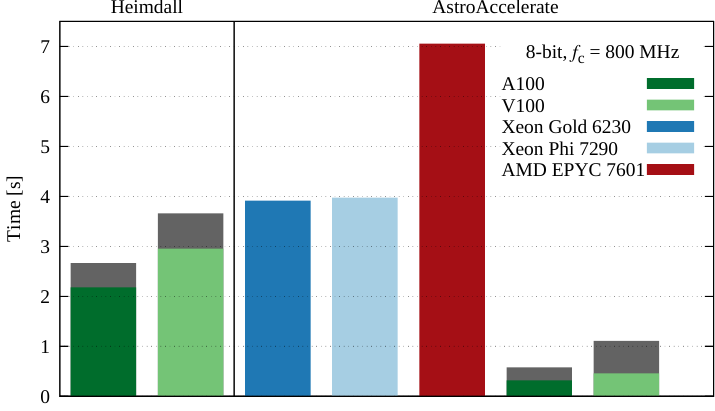}\hfill%
    \includegraphics[width=.32\linewidth]{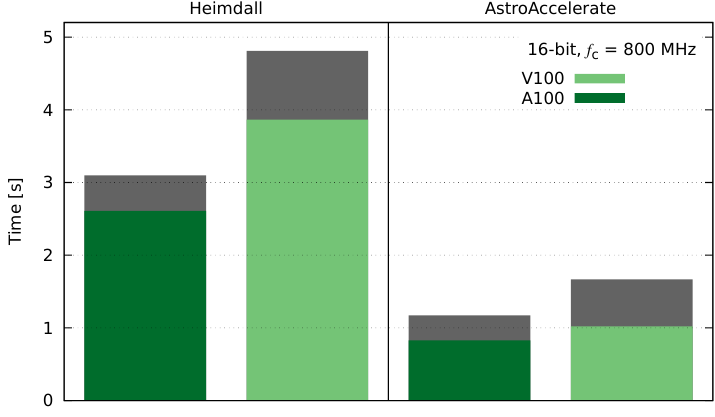}\\
    \includegraphics[width=.32\linewidth]{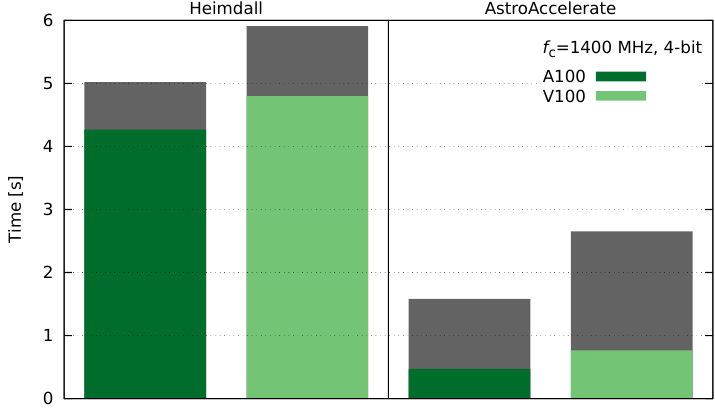}\hfill%
    \includegraphics[width=.32\linewidth]{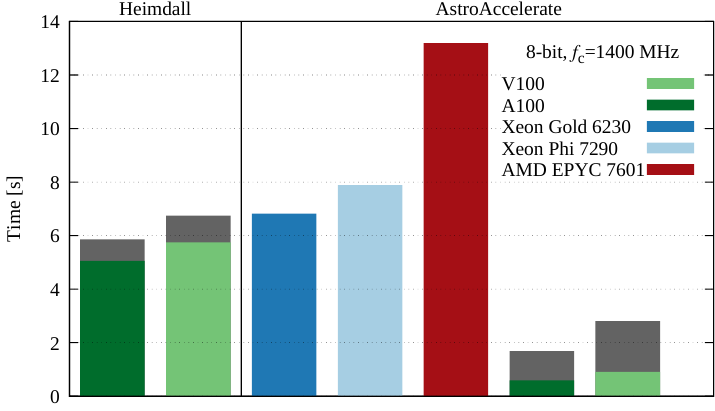}\hfill%
    \includegraphics[width=.32\linewidth]{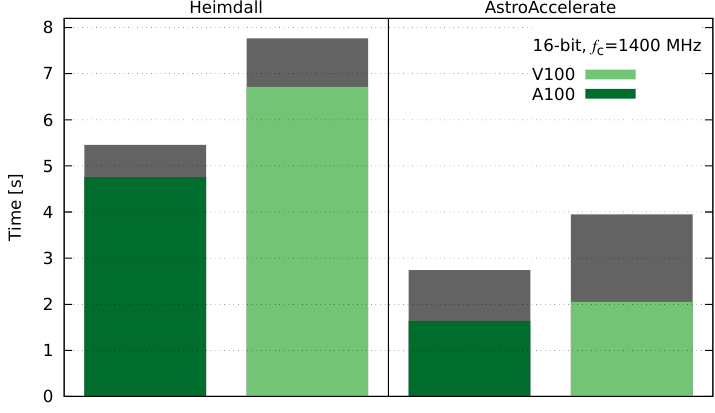}
    \caption{\label{fig:heimdall_results} Execution times of AstroAccelerate and Heimdall on different many-core platforms for three central frequencies and bit precisions operating on a simulated signal of an observation length of $\sim$300\,s for the low central frequency and $\sim$50\,s for all other central frequencies. The top row corresponds to the central frequency $f_\mathrm{c}=400$, the middle row to $f_\mathrm{c}=800$ and the bottom row to $f_\mathrm{c}=1400$\,MHz (low, mid, high). Whilst in the left column are the results for 4-bit precision, in the middle column for 8-bit precision and in the right column for 16-bit precision. The grey boxes show the execution time including all PCIe transfers and overheads needed to finish the plan. The colour (lime green, dark green) bar shows the execution time of all kernels, for example transposing of the input data, downsampling data, etc., needed by the pipeline.}
\end{figure*}
The execution times for each bit precision and many-core system (GPU or CPU) are summarised in Figure~\ref{fig:heimdall_results}. Similarly, as in the previous subsection, we found that our implementation is limited by the shared memory bandwidth (4-bit and 8-bit) or by special functional units (16-bit).

\subsection{\label{sec:output_comparison}Comparison of the de-dispersed planes}
\begin{figure}[htb!]
    \centering
    \includegraphics[width=.99\linewidth]{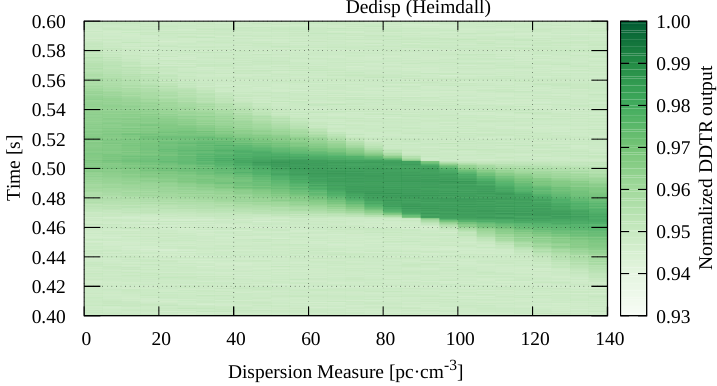}
    \includegraphics[width=.99\linewidth]{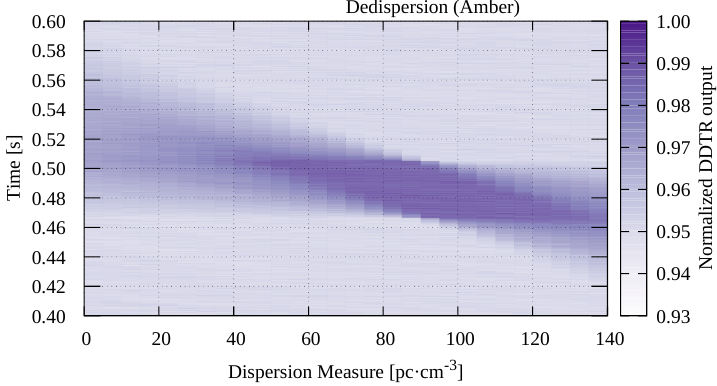}
    \includegraphics[width=.99\linewidth]{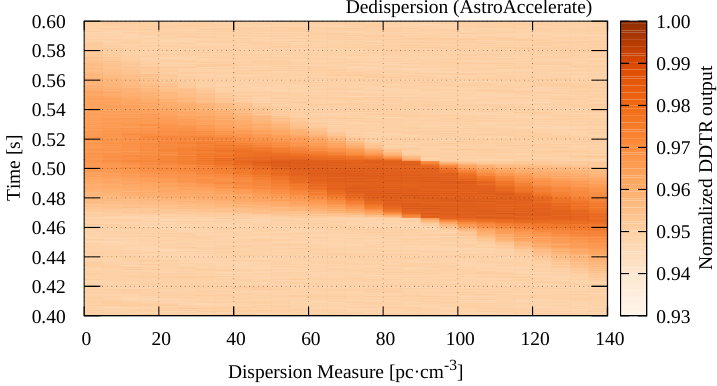}
    \caption{\label{fig:zoom_map_output} Zoomed normalised output of the dedispersion transform (DDTR)} for Heimdall (top, green), Amber (middle, purple) and AstroAccelerate (bottom, orange) of an injected signal with DM=90.0 pc\,cm$^{-3}$.
\end{figure}

\begin{figure}[htb!]
    \centering
    \includegraphics[width=.98\linewidth]{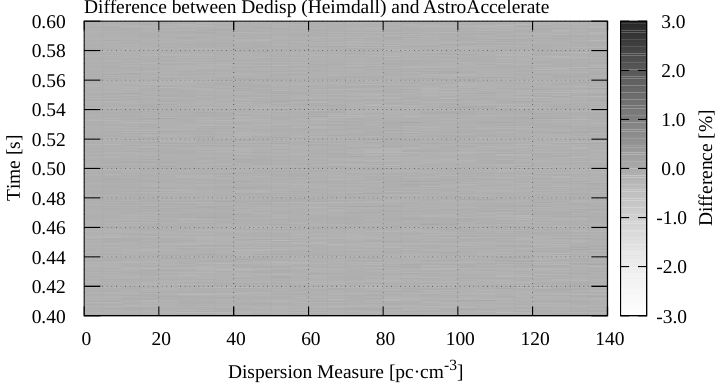}
    \includegraphics[width=.98\linewidth]{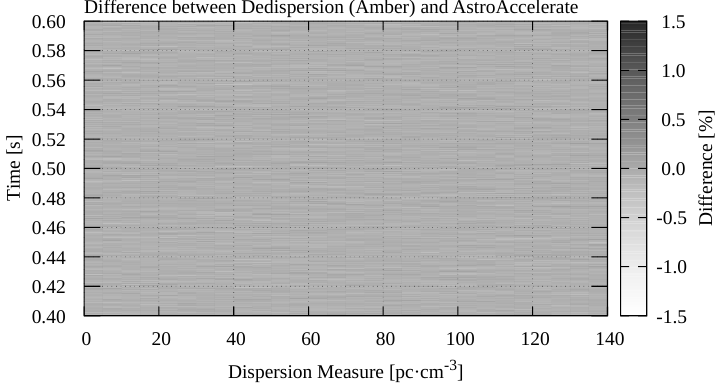}
    \caption{\label{fig:zoom_map_diff} Percentage difference of the normalised dedispersion outputs. From top to bottom: AstroAccelerate and Heimdall, AstroAccelerate and Amber.}
\end{figure}

\begin{figure}[htb!]
    \centering
    \includegraphics[width=.98\linewidth]{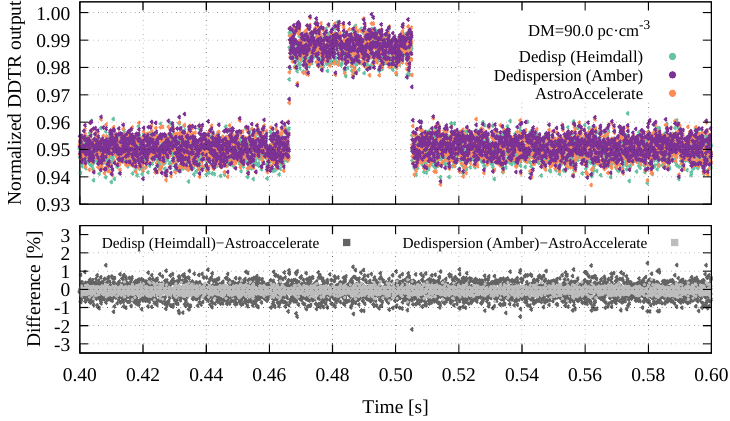}
    \includegraphics[width=.98\linewidth]{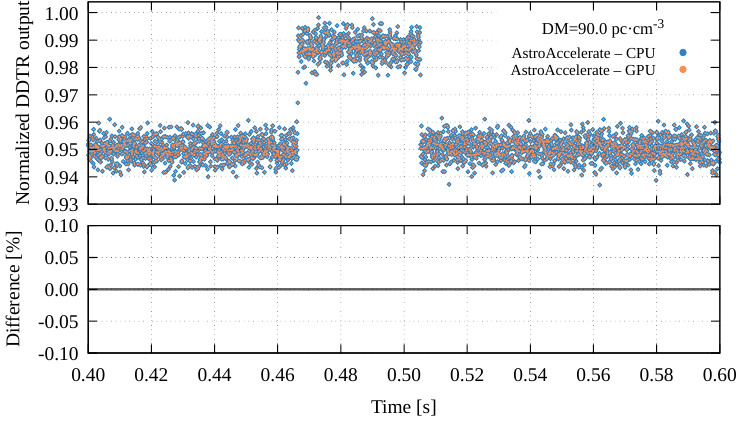}
    \caption{\label{fig:zoom_side_diff} Comparison of the dedispersed (DDTR) outputs and their percentage difference at $\mathrm{DM}=90$\,pc\,cm$^{-3}$ between AstroAccelerate, Heimdall, Amber at the top and our CPU implementation at the bottom.}
\end{figure}

The dedispersion output plane normalised to its own maximum from AstroAccelerate, Heimdall and Amber are presented in Figure~\ref{fig:zoom_map_output}. We used synthetic 8-bit data with an injected signal of DM=90.0 pc\,cm$^{-3}$. Each code was run in ``fixed'' mode with a simple survey plan searching pulses between DMs of 0--200\,pc\,cm$^{-3}$ with a step 0.5\,pc\,cm$^{-3}$. As shown in Figure~\ref{fig:zoom_map_output} all implementations recover the correct dispersion measure of the injected signal.
However, as shown in Figure~\ref{fig:zoom_map_diff}, there are small numerical differences ($\sim$1\%). These are introduced by the differences in the calculation of the time shift function, including the use of a slightly different constant of proportionality ($C_\mathrm{DM}$) used in Eq.~\ref{eq:qcp} in the different implementations. In addition, Heimdall dedispersion internally rescales the output dedispersed values, thus incurring a round-off error. Finally, in Figure~\ref{fig:zoom_side_diff}, we provide a view of a single DM trial from the dedispersed outputs between AstroAccelerate, Heimdall, and Amber and our CPU implementation around the DM of the injected signal and their percentage difference.

\subsection{Performance on selected telescopes}
\label{sec:performance_telescope}
In this section, we provide performance results for AstroAccelerate for 10 selected telescopes and their settings for two GPUs, namely the Tesla V100 GPU and A100 GPU. We have determined the DM survey plan for each telescope setup using the \texttt{DDplan.py} tool from the PRESTO pulsar search and analysis software \citep{presto:ascl:}.
We measure the performance in units of fraction of real time computed as
\begin{equation}
    R = \frac{t_\mathrm{obs}}{t_\mathrm{c}}\,,
\end{equation}
where $t_\mathrm{obs}$ is the observation time that is being processed, and $t_\mathrm{c}$ is the execution time of the AstroAccelerate pipeline. The execution time $t_\mathrm{c}$ includes the time required for dedispersion, the transpose of the input data (if necessary) and downsampling (time binning). We do not include the input data transfer time from the host to the GPU device memory. Please note that $R>1$ means that the computing is performed in real-time or greater, that is, the observation time is longer than the time needed for its processing.

Table~\ref{tab:telescopes} summarises the basic characteristics of the selected telescopes, the range of the DM survey plan and the performance of the auto-tuned AstroAccelerate software pipeline on NVIDIA Tesla V100 and NVIDIA A100 GPU in units of fraction of real-time. As we can see, in all cases, AstroAccelerate operates in real time or greater. In the worst case scenario, AstroAccelerate achieved $R=20$ on the NVIDIA Tesla V100 GPU and $R=25$ on NVIDIA A100 GPU. For more details such as the DM survey plans for individual telescopes, memory transfers, CPU performance tests and others, please see~\citet{dedispersion_data_paper}.

\begin{table*}[htb!]
\caption{\label{tab:telescopes} List of selected radio telescopes, their characteristics, search range and performance of the AstroAccelerate in fractions of real-time.}
\centering
\resizebox{\textwidth}{!}{%
\begin{tabular}{@{}llrrrrrrr@{}}
\toprule
\multicolumn{2}{c}{\multirow{2}{*}{Telescope}} &
  \multicolumn{1}{c}{\multirow{2}{*}{Central Frequency}} &
  \multicolumn{1}{c}{\multirow{2}{*}{Bandwidth}} &
  \multicolumn{1}{c}{\multirow{2}{*}{Time Sampling}} &
  \multirow{2}{*}{Channels} &
  \multicolumn{1}{c}{\multirow{2}{*}{DM range}} &
  \multicolumn{2}{c}{Fraction of Real Time} \\ \cmidrule(l){8-9} 
\multicolumn{2}{c}{} &
  \multicolumn{1}{c}{[MHz]} &
  \multicolumn{1}{c}{[MHz]} &
  \multicolumn{1}{c}{[$\mu$s]} &
  \multicolumn{1}{c}{[\#]} &
  \multicolumn{1}{c}{[pc\,cm$^{-3}$]} &
  \multicolumn{1}{r}{Tesla V100} &
  \multicolumn{1}{r}{A100} \\ \midrule
Apertif\footnote{\citet{telescope_apertif:2017}}& ALERT  & 1400  & 300  & 40.92 & 1536 & 0--10000 & 28   & 41\\ \midrule
Arecibo\footnote{\citet{Scholz20161220,Spitler20140801}}         & PALFA      & 1375  & 322       & 65.5    & 1024  & 0--9866 &    75    & 91\\ \midrule
ASKAP\footnote{\citet{Bannister20170520}}           &            & 1400  & 336       & 1265    & 336 & 0--3763 & 4613   & 5835\\ \midrule
CHIME\footnote{\citet{MIKHAILOV2018139, Chime_overview}} & FRB        & 600   & 400       & 1000    & 16384 & 0--2000 & 20     & 25\\ \midrule
Green Bank Telescope\footnote{\citet{Scholz20161220,Masui2015}}             &            & 820   & 200       & 20.48   & 512  & 0--2000 & 69     & 110\\
                &            & 2000  & 800 (600) & 10.24   & 512 & 0--1000 & 59     & 82\\ \midrule
GMRT\footnote{\citet{bhattacharyya_2017_gmrt1,Singh_2022_gmrt2, Bhattacharyya_gmrt3}}            &            & 400   & 200       & 1310.72 & 4096 & 0--2000 & 187    & 239\\ \midrule
Lovell\footnote{\citet{Scholz20161220}}          &            & 1532  & 400       & 256     & 800  &0--10000 & 622    & 837\\\midrule
Parkes - SUPERB\footnote{\citet{superb_keane:2017}} & F-pipeline & 1382  & 400       & 64      & 1024 & 0--2000 & 98     & 133\\ 
                & T-pipeline & 1382  & 400       & 64      & 1024 & 0--10000 & 89     & 127 \\ \midrule
UTMOST\footnote{\citet{bailes_utmost:2017, utmost_caleb:2017}}          &            & 835.5 & 31.25     & 655.36  & 320 & 0--10000 & 7940   & 8930\\ \midrule
VLA\footnote{\citet{vla_law:2017}}             &            & 3000  & 1024      & 5000    & 256 & 0--10000 & 254200 & 327665\\ \bottomrule
\end{tabular}
}
\end{table*}

\section{Discussion and Conclusions}
\label{sec:discussion}
In this paper, we present our CPU and GPU implementations of the incoherent dedispersion method for removing the effect of the frequency delay introduced due to the interstellar medium. Although dedispersion is only a part of the pulsar search process, its computational intensity scales rapidly with the amount of data and as such becomes a substantial contribution to the total processing time of the pipeline. We compare three different many-core implementations of the incoherent dedispersion transform, namely Dedisp by \citet{Barsdell:2012} that is part of the Heimdall pipeline, Dedispersion by \citet{Sclocco:2016} from the Amber pipeline and our implementation which is part of the AstroAccelerate project. We demonstrate that our implementation of dedispersion is faster and covers a wider range of different input data parameters.

In Figure~\ref{fig:plan_survey}, we show how the performance of AstroAccelerate depends on different input data parameters. The most important parameter for performance is the combination of sampling time and DM step. For a large enough DM step, the required data may no longer fit into the GPU's shared memory, and the cache dedispersion algorithm that relies on L1 or a slower L2 cache must be used. This leads to a performance loss of up to $4\times$. In such a case, it might be beneficial to lower the data sampling rate, for example, through time binning, and benefit from higher performance due to the smaller data size and being in a more cache-friendly regime. 

In the first benchmark in section~\ref{sec:test1} we compare all three implementations (Heimdall, Amber, our AstroAccelerate) processing input data of 4-bit, 8-bit and 16-bit precision producing a fixed number of DM trials but using different number of frequency channels. We demonstrate that the AstroAccelerate GPU implementation is at least $10\times$ faster in the case of Heimdall and from $6\times$ (with 512 channels) to $3.4\times$ (8192 channels) faster than Amber on the NVIDIA Ampere generation A100 GPU. Similar results apply for the previous generation card --~NVIDIA Tesla~V100 GPU. The worst performance compared to Heimdall is for 16-bit precision where AstroAccelerate is only $2\times$ faster.

Our CPU version of the incoherent dedispersion on all tested CPUs is comparable in performance to the Heimdall code on NVIDIA Tesla~V100 GPU for a higher number of frequency channels. Compared to the CPU version of Amber, our code is from $2\times$ (512 channels) to $15\times$ (8192 channels) faster for all tested CPUs. 

Figure~\ref{fig:flops-channel} shows that our implementations achieve, on average stable performance for all tested frequency channels in terms of FLOPS. That is, except for the case of 8192 channels for 4-bit input data due to the algorithm change. Our implementation achieves an average of ${\sim}6.5$\,TFLOPS on Tesla~V100 GPU for 4-bit, ${\sim}4.5$\,TFLOPS for 8-bit precision data and ${\sim}1.6$\,TFLOPS for 16-bit input data. Whilst on Tesla~A100 ${\sim}8.5$\,TFLOPS for 4-bit,  ${\sim}5$\,TFLOPS and 8-bit precision and ${\sim}2$\,GFLOPS in the case of 16-bit. The performance improvement of the Ampere generation compared to the Volta generation is mainly due to the increased shared memory bandwidth (from ${\sim}$14\,TB/s to ${\sim}$18\,TB/s. On the tested CPUs we get ${\sim}0.4$\,TFLOPS on KNL, ${\sim}0.33$\,TFLOPS in case of AMD EPYC 7601 and ${\sim}0.31$\,TFLOPS for Xeon Gold 6230.

The performance of other tested codes is mostly stable or improves with an increasing number of channels. Howev0er, Heimdall has a particular problem with the NVIDIA Tesla V100 GPU, as the performance decreases significantly for a high number of frequency channels, something which is not observed with the same code on the NVIDIA A100 GPU. We have observed an unusual behaviour of the Amber pipeline on the AMD CPU, where the performance decreases significantly for a higher number of frequency channels. This contradicts our expectations based on Amber's performance on the Intel CPU and the NVIDIA GPUs. This may indicate the use of platform-specific optimisation in Amber or the lack of OpenCL support for AMD EPYC CPU. It also shows that even though the OpenCL programming model is easily portable to different many-core systems, it does not always guarantee good performance. 

In Section~\ref{sec:test2}, we perform a pipeline test for three different central frequencies ($f_c=$400 (low), $f_c=$800 (mid) and $f_c=$1400 (high)) searches running from 0 up to 1500, 2000 and 3000\,pc\,cm$^{-3}$ respectively with the sampling rate and the number of channels depending on central frequency. We execute all benchmarks for AstroAccelerate and Heimdall for 4-bit, 8-bit and 16-bit precision input data (where applicable) on the GPUs as well as on the CPUs. We find that both AstroAccelerate and Heimdall on all tested platforms operate in real-time regimes. That is, the end-to-end execution time of a pipeline (that includes all required operations like time binning, data transfers to the GPU memory, dedispersion, etc.) for the selected dedispersion plan is lower than the observation time of the input data. Overall, our GPU version, compared to Heimdall, is 4--8x faster for all tested central frequencies and input data precisions.

The performance of our CPU implementation is comparable with Heimdall running on either NVIDIA V100 GPU or NVIDIA A100 GPU. Only at mid-central frequencies is Heimdall substantially faster. Depending on the structure of the pipeline, the CPU dedispersion code may offer a way to distribute the tasks between the CPU and GPU, where the GPU can, for example, perform FDAS~\citep{Ransom2002:FDAS, Sofia:2018:FDAS} or JERK search~\citep{Andersen2018:JERK, Adamek2020:JERK} while the CPU calculates DM trials, thus enabling heterogeneous systems.

Lastly, we run AstroAccelerate on several telescope setups with different survey plans. The plans are created with the \texttt{DDplan.py} tool from PRESTO up to the typical dispersion measures ranges given in the corresponding telescope articles. On both NVIDIA V100 GPU and NVIDIA A100 GPU, we achieve real-time performance in all cases, i.e., from 20--5000 fraction of real-time and even 200\,000 fraction of real-time for the VLA telescope. 

\section*{Acknowledgements}
%
The authors would like to thank Aris Karastergiou, Steve Roberts, Ben Stappers, Mitch Mickaliger, Duncan Lorimer, Maciej Serylak, Evan Keen and Cees Bassa for useful comments, advice and help with testing. The project has also received support from the members of the Oxford pulsar group, Christopher Williams, Griffin Foster and Jayanth Chennamangalam, as well as support from the Time Domain Team, a collaboration between Oxford, Manchester and MPIfR Bonn, to design and build the SKA pulsar search capabilities. The authors would like to express their gratitude to the Research Centre for Theoretical Physics and Astrophysics, Institute of Physics, Silesian University in Opava for institutional support.

This work has been supported by the Institute for the Future of Computing, Oxford Martin School and STFC grants ST/N003713/1, ST/P005446/1, ST/R000557/1, ST/W001969/1 and Silesian University in Opava IGS/11/2022 grant. Also, the authors would like to acknowledge the use of the University of Oxford Advanced Research Computing \citep{Richards:2015:ARC} facility in carrying out this work.
\label{lastpage}
\bibliographystyle{aasjournal}
\bibliography{dedispersion_arxiv}
\end{document}